# Carbon fibre tips for scanning probe microscopy based on quartz tuning fork force sensors

A Castellanos-Gomez[1], N Agraït[1,2,3] and G Rubio-Bollinger[1,2]

[1] Departamento de Física de la Materia Condensada (C–III).
Universidad Autónoma de Madrid, Campus de Cantoblanco, 28049 Madrid, Spain.
[2] Instituto Universitario de Ciencia de Materiales "Nicolás Cabrera".
[3] Instituto Madrileño de Estudios Avanzados en Nanociencia
IMDEA-Nanociencia, 28049 Madrid, Spain

E-mail:
    gabino.rubio@uam.es
    andres.castellanos@uam.es

We report the fabrication and the characterization of carbon fibre tips for their use in combined scanning tunnelling and force microscopy based on piezoelectric quartz tuning fork force sensors. We find that the use of carbon fibre tips results in a minimum impact on the dynamics of quartz tuning fork force sensors yielding a high quality factor and consequently a high force gradient sensitivity. This high force sensitivity in combination with high electrical conductivity and oxidation resistance of carbon fibre tips make them very convenient for combined and simultaneous scanning tunnelling microscopy and atomic force microscopy measurements. Interestingly, these tips are quite robust against occasionally occurring tip crashes. An electrochemical fabrication procedure to etch the tips is presented that produces a sub-100 nm apex radius in a reproducible way which can yield high resolution images.

## 1. Introduction

The techniques of scanning tunnelling microscopy (STM) [1] and atomic force microscopy (AFM) [2] have proven to be extremely useful tools in the exploration of properties of matter at the nanoscale. Both techniques make use of sharp tips to probe properties at the atomic scale, with reliability, resolution and stability strongly dependent on the tip's physical and chemical properties. With these factors in mind, experimental efforts to improve the quality of the fabricated tips [3-5] and to develop tips based on new materials [6-8] have been motivated by the desire to gain improvement over previous methods. Recently, combined STM/AFM microscopes based on quartz tuning fork force sensors (tuning fork combined STM/AFMs hereafter) have been developed [9, 10]. These sensors are fabricated by attaching a tip to one prong of a miniaturized quartz tuning fork. Tuning fork force sensors are stiffer than conventional microfabricated cantilevers, preventing the tip jumping to contact at very small tip to sample distances even with small, or zero, oscillation amplitudes [11, 12]. This fact makes tuning fork force sensors capable of obtaining simultaneously force and tunnelling current maps and hence being more convenient for combined STM/AFMs than conventional cantilevers. Moreover tuning fork force sensors can have a much higher quality factor $Q$ than cantilevers. The high $Q$ factor in combination with a stable small oscillation amplitude operation makes tuning fork force sensors very sensitive to atomic scale forces without compromising the STM operation of the microscope [13, 14]. A further advantage of these microscopes is that they can be implemented under extreme environments such as ultrahigh vacuum (UHV), low temperatures and high magnetic fields making them a versatile tool for studying in detail electronic and mechanical properties of surfaces. In these combined microscopes the tip material has to be chosen carefully in order to ensure an optimal operation in both STM and AFM schemes. Although platinum/iridium (PtIr) and tungsten (W) are the most commonly used STM tips, their use in tuning fork combined STM/AFM can strongly modify the tuning fork dynamics degrading the $Q$ factor and the force sensitivity of the tuning fork sensor [15-18]. It is, therefore, desirable that tips have a low impact on the tuning fork dynamics whilst keeping a high electrical conductance. Carbon based tips [19-25] are promising tips for combined STM/AFMs due to their electrical and mechanical properties. The capability of chemically modifying the tip surface [26] may open up new possibilities to study molecular electronic devices or biofunctionalized surfaces. The most extended carbon-based tips are the carbon nanotube tips [21-23, 27-30] and the electron beam deposited amorphous carbon tips [24, 25]. These tips have, however, not been widely used in combined STM/AFMs possibly due to the need for specialised tools for their fabrication and testing [21]. Carbon fibre tips, whilst sharing many of the remarkable mechanical, electrical and chemical properties of other carbon-based tip, are easy to fabricate and handle. This makes them promising candidates for use in a combined STM/AFM.

Here we fabricate carbon fibre tips for their use in such a combined tuning fork STM/AFM. We have developed a procedure for electrochemically etching carbon fibre tips to increase their aspect ratio and to achieve better spatial resolution in STM/AFM images. The STM and AFM imaging capabilities of carbon fibre tips have been tested using a homebuilt combined tuning fork STM/AFM. Interestingly these tips have been found to be quite robust against occasional tip crashes, maintaining high lateral resolution images even after several such events.

## 2. Properties of Carbon fibre tips

For simultaneous tunnelling current and force measurements the tip should meet specific requirements for both its electrical conductance and mechanical properties. For STM operation a tip with high resistance to oxidation and a high electrical conductance is desirable. The use of carbon fibre tips is therefore motivated by their high oxidation resistance [20, 31]. Indeed our carbon fibre tips have been used in air for several weeks without significant degradation in their STM operation. We have measured the electrical resistivity $\rho$ of the carbon fibres used in this work[1] ( $\rho = (1.3 \pm 0.3) \times 10^{-5}$ $\Omega$m ) and for typical tip dimensions[2] the series resistance is $50$ $\Omega$.

With a conventional STM tip (PtIr or W) attached to one of the prongs of a tuning fork force sensor the *Q* factor and force sensitivity of the AFM lowers due to the imbalance between tuning fork prongs [15-18] which changes the tuning fork dynamics. This effect can be minimized by using low mass, rigid tips. Carbon fibre tips can be very lightweight because of the combination of their low density $\rho$ (around 1.6 gr cm$^{-3}$) and small fibre diameters (between 5 – 10 µm). On the other hand PtIr or W tips are more dense, their density being around 19 gr cm$^{-3}$ with wire diameters typically in the region of 10 – 50 µm. Figure 1 shows how the *Q* factor of a tuning fork loaded with a carbon fibre tip is only 2% lower than that of a tipless tuning fork. The reduction in the *Q* factor is about 40% with a PtIr tip attached similar to those used in refs [32, 33]. In addition we find that carbon fibre is easier to handle than thin metallic wires due to their straightness over lengths in the range of centimetres, metallic wires tending to curl.

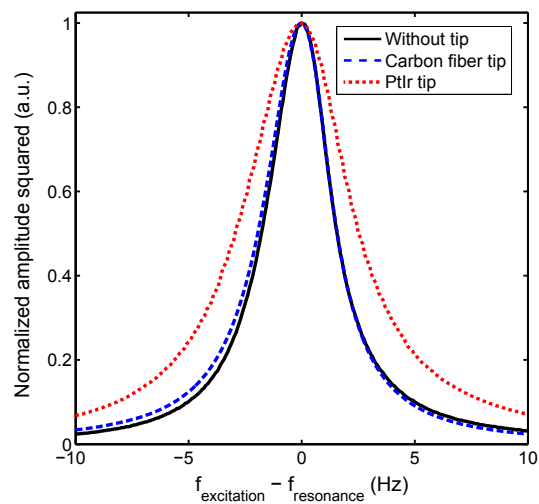

Figure 1.(Colour online) Resonance spectra of three quartz tuning forks[3]: (solid black line) without tip, (dashed blue line) with a tip made of a 7 µm diameter 300 µm long carbon fibre and (red circles) with a tip made of a 25 µm diameter 300 µm long PtIr wire. Their resonance frequencies $f_0$ are 32.768 KHz, 32.610 KHz and 31.907 KHz respectively. Their quality factors *Q* are 9712, 9446 and 5825 respectively.

To increase the stability of combined STM/AFMs it is desirable to use tips with high resonance frequencies which are less sensitive to external mechanical perturbations. A high ratio between the Young's modulus and the density of the tip's material guarantees a high resonance frequency of the tip. Carbon fibre tips meet these requirements due to their high Young's modulus (between 200 – 800 GPa depending on the fibre used) and their low density. We have measured the Young's modulus *E* of our carbon fibre tips ( $E = 280 \pm 20$ GPa ). To determine the Young's modulus we fabricate carbon fibre cantilevers and measure the change in resonance frequency when small test masses are attached at their free end. Then we fit the resulting dependence to the one expected for cylindrical cross-section cantilevers. Other important characteristic of the tips used in combined STM/AFMs is their mechanical resistance to crashes with the surface. In order to get tips as durable as possible, tip materials with high yield strength have to be considered. These materials are capable of standing large strains without suffering plastic deformation, recovering their original shape when the stress is removed. The yield strength of carbon fibres is around 4 GPa while for PtIr or W is about 0.5 GPa, making carbon fibre tips better candidates for durable tips. Furthermore, in most available carbon fibres the yield point and the rupture occur at the same strain which means that the deformation is elastic up to rupture [34, 35].

---

[1] PAN based carbon fibre manufactured by Hercules inc.: Reference: AS4-12K. Single fibres can be easily extracted from the carbon fibre rope provided by the manufacturer.
[2] Protruding length $\leq 150$ µm and $7$ µm in diameter.
[3] Tuning forks manufactured by Rakon. Reference: XTAL002997.

## 3. Electrochemical etching

Although in STM the use of mechanically fabricated tips (by simply cutting a metallic wire) is rather common, the AFM resolution strongly depends on the tip sharpness because of the presence of long range interactions between the tip and the sample. In this work we have developed an electrochemical procedure to etch carbon fibre tips.

The setup used to etch electrochemically the carbon fibres is very similar to the one used to etch metallic tips. A 5-10 mm long fibre is extracted from the fibre rope. One end of the fibre is immersed a few microns into a drop of 4M KOH solution suspended in a 4 mm inner diameter gold ring[4]. A voltage bias difference of 5-6 Volts is applied between the unimmersed fibre end and the gold ring which is grounded. The etching takes place over a period of tens of seconds until the fibre breaks, opening the electrical circuit and stopping the etching. Afterwards the fibre is rinsed with distilled water. Reproducible tips with sub-100 nm apex radius can be obtained following this procedure as shown in figure 2. A three axis micrometer manipulator and a stereoscopic microscope are used to mount and glue the tip overhanging from the free end of one of the tuning fork prongs. The rest of the fibre can be easily cleaved after the glue is cured by bending until fracture. We use silver conductive epoxy[5] to connect electrically the carbon fibre to one of the quartz tuning fork electrodes which is used as the STM voltage bias. In this way only one extra electrical connection is needed to implement a tuning fork sensor on an STM.

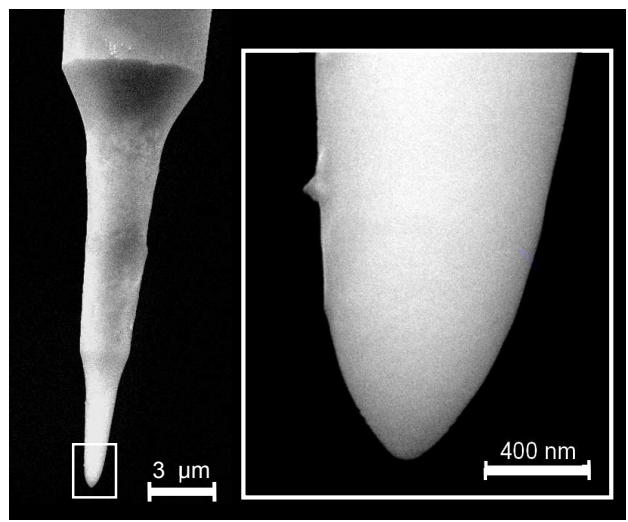

---

[4] Made from a 125 µm diameter gold wire.
[5] Silver loaded conductive epoxy purchased from RS-Online. RS number: 186-3616.

Figure 2. Scanning electron micrograph of a carbon fibre tip electrochemically etched following the procedure described above. The curvature radius of the tip apex is estimated to be 55 nm.

**4. Experimental results**

In the following we will characterize the behaviour of these carbon fibre based tips for combined STM/AFM operation with a home-built microscope [32]. First a gold[6] (111) surface was studied to characterize the capabilities of simultaneous STM/AFM operation, tunnelling spectroscopy and electrostatic force microscopy in order to probe their operation in STM and AFM modes. Afterwards the lateral resolution achieved in STM and AFM modes was studied using several samples. All the measurements were carried out in an air environment.

*4.1. STM/AFM capabilities*

We have studied the interaction force between a carbon fibre tip and a Au (111) surface when the tip is approached to the surface until reaching the tunnelling regime. A frequency modulation mode (FM AFM) has been used [36] in which the tuning fork is driven at its resonance frequency using a *phase locked loop* (PLL). An attractive (repulsive) force gradient acting between tip and sample produces a positive (negative) shift on the resonance frequency (top panel in figure 3. In the limit of a small oscillation amplitude the force gradient can be easily related to the frequency shift [15, 37] and the force *vs.* distance curve can be obtained by integration. An oscillation amplitude of 0.2 nm$_{RMS}$ has been used in this measurement to guarantee the validity of the small oscillation amplitude limit. The time averaged tunnel current during the oscillation of the tip is measured simultaneously with the frequency shift (bottom panel in figure 3) [13, 38, 39]. We find that tunnel currents of up to 100 pA can be obtained while in the attractive force regime, that is in the non-contact regime. The tunnel current *vs.* tip-sample distance shows an exponential dependence with two different decay constants (inset in figure 3). The tunnel barrier height $\phi$ deduced from the exponential dependence is 0.8 eV for the lower tunnelling currents. Notice that tunnel barrier height values lower than 1 eV are common in STM operation in air [40] and are attributed to the three-dimensional tunnelling through an interfacial layer of water present in the tunnel junction [41]. We observe a sudden drop of the barrier height from 0.8 eV to 0.03 eV where the force changes sign, i.e. when the tip and the sample are in mechanical contact. This result is in good agreement with previous works in which it is suggested a drastic lowering of the barrier height due to the surface deformation induced by the strong repulsive forces during the tip-sample contact [41, 42]. The electrical conduction in the repulsive regime has been attributed to tunnelling through an insulating barrier [41] formed by oxides or adsorbates on the surface when exposed to air. In this situation when there is mechanical contact with the adsorbate layer electrons can tunnel through the adsorbate barrier. We have also measured simultaneously the change of the *Q* factor of the tuning fork during the approach (bottom panel in figure 3). The *Q* factor falls by 40% before entering the tunnel regime, and can be attributed to several sources such as ohmic dissipation [43, 44] or force gradient induced imbalance of tuning fork prongs [15-18].

---

[6] Arrandee 11x11 mm$^2$ gold substrate. It has been flame annealed to obtain clean atomically flat terraces following the Au (111) orientation.

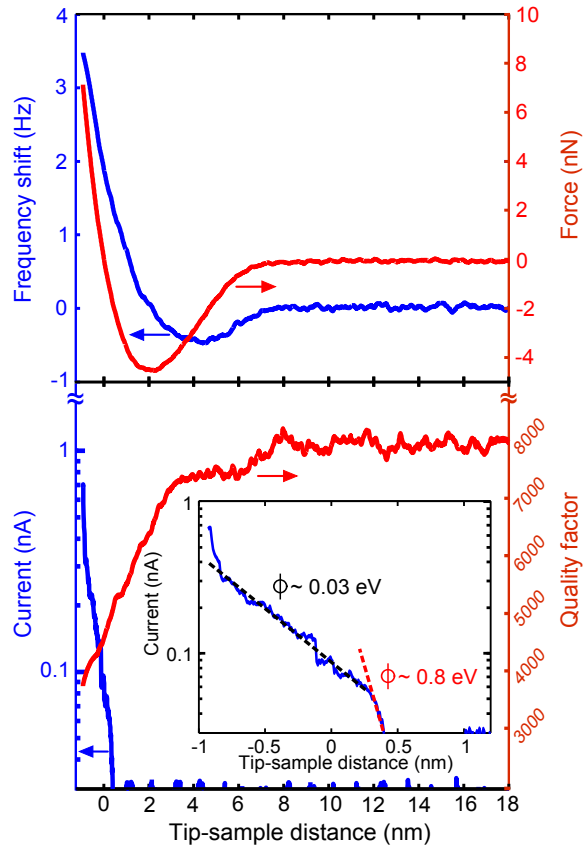

Figure 3. (Colour online) Simultaneous measurement of the tuning fork frequency shift (top panel, left axis) and of the tunnel current (bottom panel, left axis) during the tip-sample approach. The small oscillation amplitude model (A = 0.18 $nm_{RMS}$) is appropriate because the force decay length (~2 nm) is much larger than the tip oscillation amplitude (0.18 $nm_{RMS}$). (Top panel, right axis) Force *vs.* distance curve. (Bottom panel, right axis) Change in the *Q* factor of the tuning fork measured also simultaneously. (Bottom panel inset) Detail of the tunnel current *vs.* distance. Parameters: $k_{prong}$ ~ 12000 N/m; ; $f_0$ = 31775 Hz; *Q* = 8295 ; A = 0.18 $nm_{RMS}$; $V_{bias}$ = 10 mV. Notice that we have chosen as origin in the tip-sample distance axis the point in where the integrated interaction force changes its sign.

### 4.2. Tunnelling spectroscopy capabilities

The electronic local density of states (LDOS) can be probed with an STM by means of the tunnelling spectroscopy technique, in which the tip-sample bias voltage dependence of the tunnel conductance is measured [45]. We have carried out spectroscopic measurements using a carbon fibre tip and a Au (111) surface. Figure 4 shows the average of 32 tunnel current *vs.* bias voltage (IV) curves (left axis). The average tunnel differential conductance *vs.* bias voltage (dI/dV) has been obtained by numerical derivation of 32 *IV* curves (right axis in figure 4). The parabolic shape of the dI/dV curve is very similar to the one obtained using metallic tips on a highly ordered pyrolitic graphite (HOPG) surface [46].

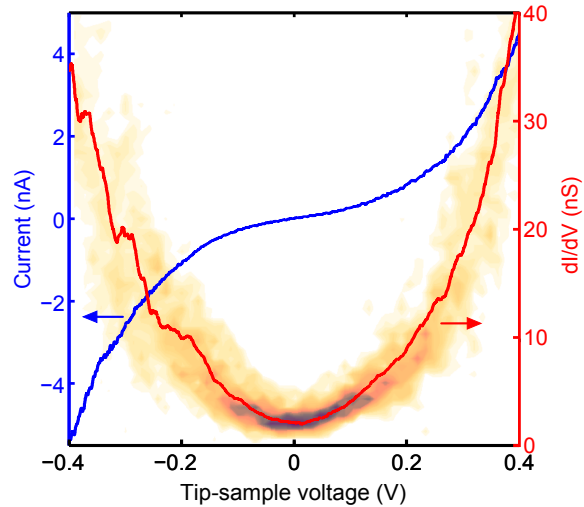

Figure 4. (Colour online) Average tunnelling current *vs.* voltage curve (left axis) obtained from 32 individual curves measured on an Au (111) surface and the corresponding differential conductance curve (right axis). The density plot obtained from the 32 individual dI/dV curves indicates the dispersion around the average differential conductance curve. The darker (lighter) the zone, the lower (higher) dispersion in the differential conductance value.

### *4.3. Electrostatic force microscopy capabilities*

If the AFM tip and the sample are both conductive materials the AFM can be employed to probe electrostatic forces. When the tip and sample are close enough they form a nanocapacitor. The electrostatic force gradient between the plates of this nanocapacitor depends quadratically on the bias voltage applied between them [47]. This force gradient reaches a minimum when the applied voltage counteracts the tip-sample work function difference ($\Phi_{sample} - \Phi_{tip}$), i.e. the contact potential difference $V_{CPD}$. Figure 5 shows the frequency shift as a function of the bias voltage applied between the tip and a Au (111) surface. A $V_{CPD}$ of 150 mV is obtained from a quadratic fitting of the experimental data. This value is in agreement with the $V_{CPD}$ of 100 – 200 mV previously reported in the literature for gold islands grown onto HOPG [47] and with the $V_{CPD}$ of 130 mV measured with a gold tip onto an HOPG surface [48].

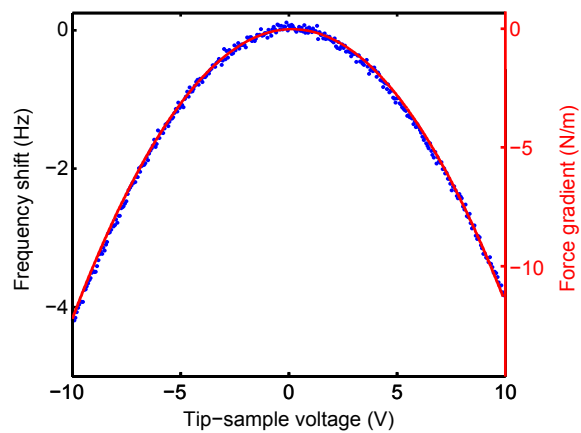

Figure 5. (Colour online) Tuning fork frequency shift as a function of the tip-sample voltage difference. A contact potential difference of 150 mV is obtained from the position of the parabola vertex. Parameters: $k_{prong}$ ~ 12000 N/m; ; $f_0$ = 31775 Hz; $Q$ = 8295 ; A = 0.56 nm$_{RMS}$; tip-sample distance ~ 2.1 nm.

### *4.4. STM imaging*

We have used a freshly cleaved surface of HOPG to probe the spatial resolution of carbon fibre tips in STM operation. Figure 6 shows a constant height STM image of an atomically flat terrace of HOPG. The characteristic triangular (figure 6.a) and honeycomb (figure 6.b) lattices can be observed by adjusting the tip-sample distance. The observation of both lattices can be attributed to different interaction of the surface graphite plane with consecutive planes [49, 50]. Due to the AB stacking of graphite there are two types of carbon atom sites, A and B. In the B-sites the carbon atoms are located over the centre of the hexagon of the underlying layer while in the A-sites the carbon atoms are over another carbon atom of the layer below the surface. Carbon atoms in B-sites have a higher contribution to the tunnel current than the atoms in A-sites which explains the triangular lattice observed in STM measurements (see figure 6.a and 6.c) [51]. Nevertheless tip-sample interactions can reduce the coupling of the surface layer with the layer underneath in which case all atoms in the surface equally contribute to the tunnel current. In this situation both atoms of the lattice can be resolved (see the profile in figure 6.c) and the honeycomb lattice can be measured by STM (figure 6.a). Although atomic resolution images of HOPG surfaces had been achieved using other carbon-based tips [19, 20, 52], the observation of both the triangular and the honeycomb lattices using a carbon-based tip had not been reported.

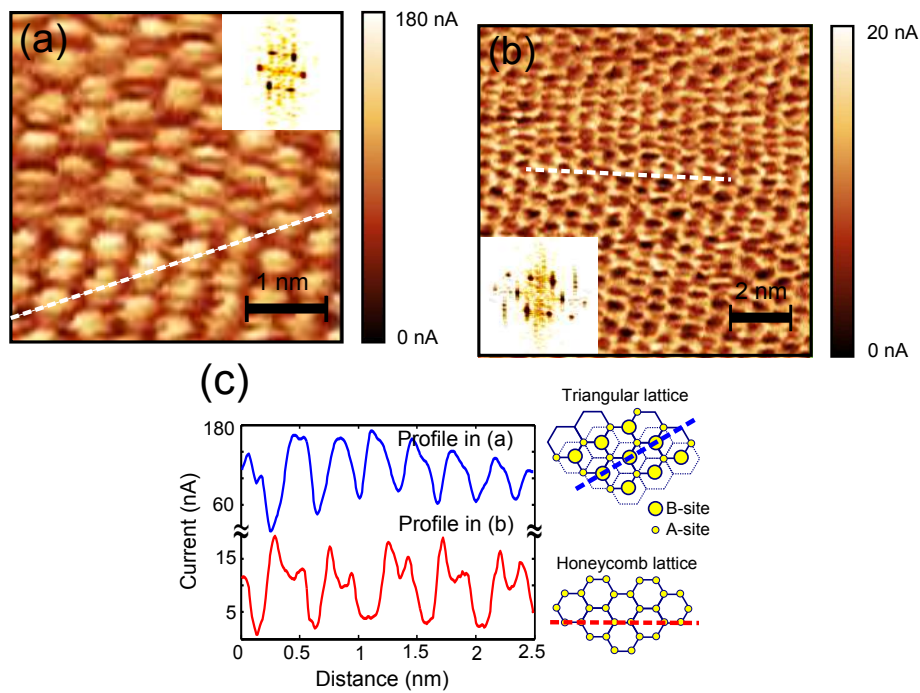

Figure 6. (Colour online) Tunnel current image in constant height STM mode of a freshly cleaved surface of HOPG exhibiting both triangular (a) and honeycomb (b) lattices, their fast Fourier transform is also shown (insets). Profiles following the dashed lines in (a) and (b) are shown in (c). A cartoon in (c) illustrates that the difference in the images and the profiles is induced by the interaction of the surface layer with the underlying layer (see the text). Parameters: $V_{bias}$ = 100 mV.

*4.5. AFM imaging*

A 30 nm thick gold film thermally evaporated onto a silicon oxide substrate has been used to probe the lateral resolution and the stability of the combined STM/AFM microscope using electrochemically etched carbon fibre tips. We selected this sample because its highly corrugated grain structure cannot be resolved using a blunt tip. Figure 7a shows an AFM topography image of the Au thin film measured in FM-AFM mode. Due to the absence of long-range ordering in the morphology of the sample it is convenient to use the circularly-averaged autocorrelation function *g(r)* [53] instead of the 2D Fourier transform to statistically analyze the topographic data (figure 7b) as in previous studies of rough surfaces [53-55]. A mean value for the radius of the gold grains of 18 nm is obtained from the position of the first zero-crossing of *g(r)* [53, 54]. The mean distance between gold grains is 70 nm, determined from the position of the first maximum of *g(r)* [55]. Individual gold grains can be spatially resolved due to the sharpness and the high aspect ratio of the tip. Using the etching recipe introduced in section 3 we routinely obtain similar results in these samples. It is remarkable that even at room conditions our spatial resolution is compatible with the one obtained by Rychen *et al.* [56] where 30 nm gold grains were resolved on a gold film evaporated on glass using a combined STM/AFM operated at low temperature (T = 2.5K) with an electrochemically etched W tip. Note that at low temperature the tuning fork *Q* factor is much higher than at room conditions [57], thermal drift is less severe and the presence of contamination gases is reduced by the cryogenic vacuum.

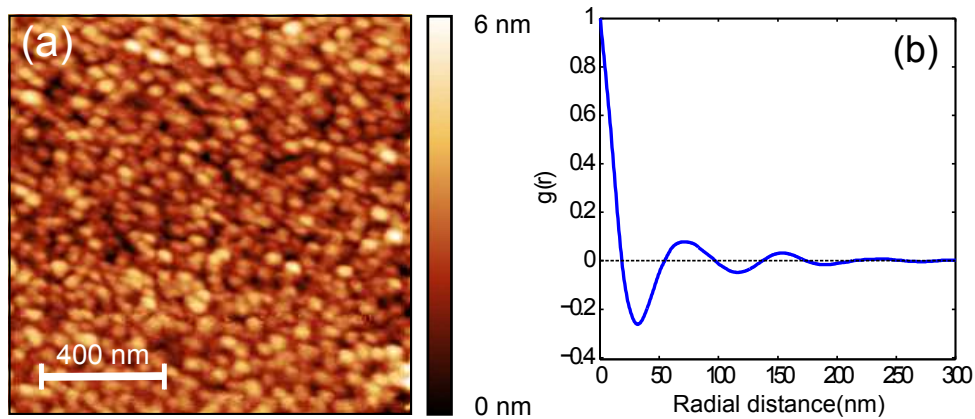

Figure 7. (Colour online) (a) Frequency modulation AFM topography of a rough gold surface. Parameters: $k_{prong}$ ~ 12000 N/m; $f_0$ = 32300 Hz; $Q$ = 8980; $\Delta f$ = 2 Hz; $A$ = 0.28 $nm_{RMS}$; $V_{bias}$ = 10 mV. (b) Circularly averaged autocorrelation function *g(r)* corresponding to the topography image in (a).

*4.6. Combined STM/AFM imaging*

The same Au thin film was also imaged in dynamic STM mode [12, 33] (figure 8). In this technique the tip is oscillated with a small amplitude (0.3 $nm_{RMS}$) and close enough to the surface to establish a tunnel current. The time averaged tunnel current is used as feedback signal to keep the tip-sample separation constant and thus to get the surface topography (figure 8a). Resolution in STM images is usually higher than in AFM because of the exponential dependence of the tunnel current with the tip-sample distance and because of the closer proximity between tip and sample during the STM operation. In figure 8a gold grains are better defined than in figure 7a making possible to verify the values obtained in previous section for the mean radius of the gold grains and the mean distance between grains. The mean grain radius and mean grain distance obtained from the circularly-averaged autocorrelation function in figure 8c are 19 nm and 65 nm respectively which are compatible with the values obtained in previous section from the AFM topography of the same sample. Simultaneously the tuning fork frequency shift is measured (figure 8b) as in FM-AFM mode. In this way the tip-sample interaction can be studied during an STM scan. Although in constant current STM operation the tip-sample distance is kept constant, the frequency shift image shows contrast. The origin of this contrast is related to the different tip-sample distance dependence of the tunnel current and the frequency shift. The fact that the frequency shift is also sensitive to long range interactions suggests that there can be differences in the force gradient for rough surfaces even if the tunnelling distance is kept constant. While the quantitative relation between tunnelling current and force gradient is highly interesting [13] it would require an additional detailed analysis.

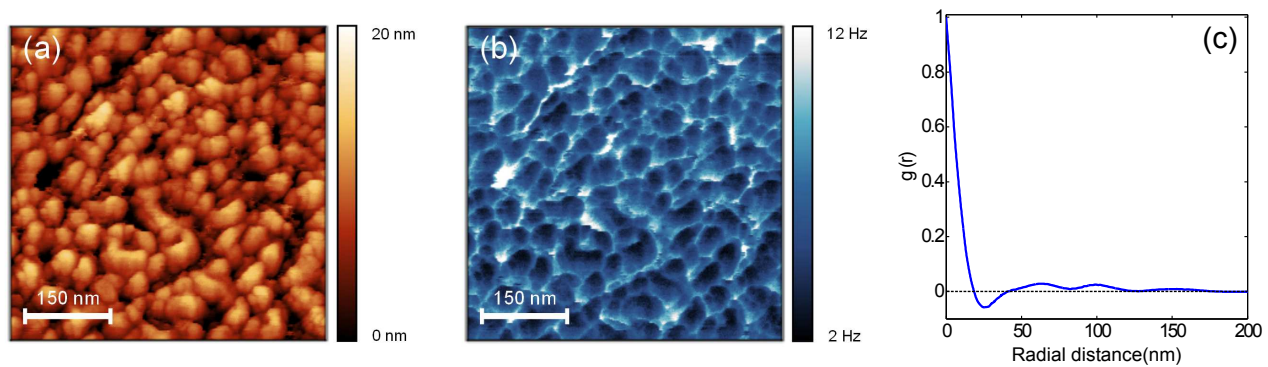

Figure 8 (Colour online) (a) Dynamic STM topography of a rough gold thin film. (b) Simultaneously measured resonance frequency shift. Parameters: $\langle \Delta f \rangle \sim 5$ Hz; $A = 0.31$ nm$_{RMS}$; $\langle I \rangle = 1$nA; $V_{bias} = 100$mV. (c) Circularly averaged autocorrelation function $g(r)$ obtained from the topography in (a).

### 4.7. Unetched tips behaviour

As shown in section 3 the long range interactions between tip and sample necessitate the use of sharpened tips to ensure high spatial resolution in AFM mode. Surprisingly we have found that even unetched carbon fibre tips (simply cut with scissors) operate in a reproducible way in STM and AFM modes. Due to the negligible presence of plastic deformations in carbon fibres, they are elastically deformed until rupture by the action of the scissors. This process can produce very sharp asperities which can be used as tips in combined STM/AFM measurements. High spatial resolution without tip artefacts can be achieved using unetched tips when atomically flat surfaces are imaged. Figure 9 shows an example of an AFM topography image of monoatomic steps on HOPG using an unetched carbon fibre tip. Nevertheless better spatial resolution can be achieved on rough surfaces with electrochemically etched tips.

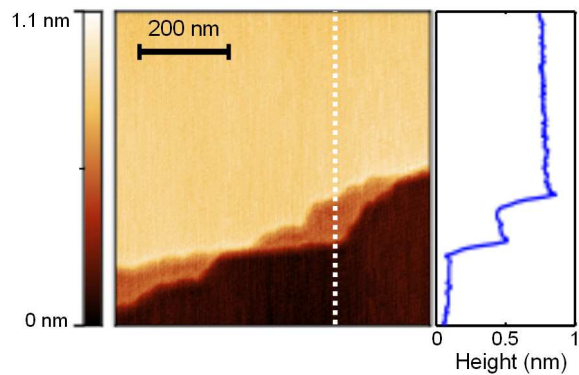

Figure 9. (Colour online) Frequency modulation AFM topography of HOPG monoatomic steps using an unetched carbon fibre tip. A profile across the dashed line is shown in the right panel. Parameters: $k_{prong}$ ~ 12000 N/m; ; $f_0$ = 32450 Hz; $Q$ = 9120 ; $\Delta f$ = 2 Hz; A = 0.7 $nm_{RMS}$.

Unetched carbon fibre tips have in addition a high mechanical resistance to occasional tip crashes. Figure 10 (top) shows the AFM topography of a titanium/gold structure lithographed on silicon oxide before (left) and after (right) crashing a carbon fibre tip intentionally (200 Hz frequency shift). The same procedure has been carried out with an unetched PtIr tip for comparison (figure 10 bottom). This test has been repeated under the same conditions using more than 10 tips giving similar results to the ones shown in figure 10. The effect of the crash on image resolution is much stronger for the PtIr tip. This effect can be attributed to irreversible plastic deformation of the PtIr tip apex. On the other hand, the carbon fibre tip doesn't suffer such strong deformations probably due to the high yield strength of the carbon fibre that prevents any plastic deformations. Etched carbon fibre tips are also very resistant to crashes but the spatial resolution after a hard crash becomes similar to the one achieved when unetched carbon fibres are used indicating a clean fracture of the tip apex. This observation is also in agreement with the absence of plastic deformations on carbon fibres. This characteristic makes carbon fibre tips very useful for tuning fork STM/AFMs designed to operate in low temperature or UHV environments in which changing the tip can be time consuming.

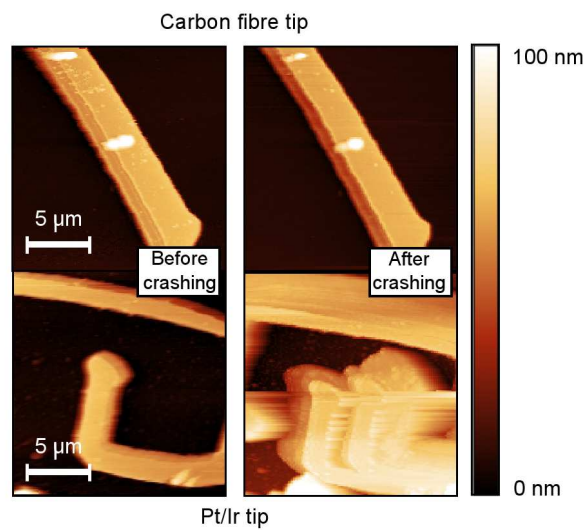

Figure 10: (Colour online) AFM topography of lithographed titanium/gold structures on silicon oxide. On the left (right) the topography before (after) indenting the tip on the silicon oxide until reaching 200 Hz frequency shift. On the top (bottom) are shown the results obtained by using a carbon fibre (PtIr) tip.

## 5. Conclusions

We have fabricated and characterized carbon fibre tips for their use in combined STM/AFMs based on quartz tuning fork force sensors. This is the first time that single carbon fibre tips have been used in scanning probe microscopy. We find that the use of carbon fibre tips results in a lower impact on the dynamics of quartz tuning fork sensors than conventional metallic tips, yielding high *Q* factor and force sensitivity. Carbon fibre tips also exhibit higher resistance to occasional tip crashes than metallic tips, which are more prone to plastic deformations. In addition, carbon fibre tips have high electrical conductivity and oxidation resistance which makes them very suitable for combined STM/AFM measurements and may also open up new possibilities for studying local properties of surfaces with scanning probe techniques under extremely oxidative environments, measuring electric transport through carbon-based nanostructures and probing mechanical and electrical properties of molecules on surfaces. The chemical nature of the tip can play an important role in a broad range of experiments involving chemical bonding with carbon atoms of the tip such as those related to biofunctionalization and single-molecule electronics. We present an electrochemical procedure to etch carbon fibre tips which produces sub-100 nm tip apex radius in a reproducible way increasing the lateral resolution in AFM measurements. We conclude that carbon fibre tips mounted on quartz tuning fork force sensors can be reliably used in force and/or tunnel current *vs.* distance measurements, electron tunnelling spectroscopy, electrostatic force microscopy, AFM, atomic-resolution STM and simultaneous STM/AFM microscopy.

## 6. Acknowledgements

The authors wish to thank P. Joyez for the scanning electron microscopy images of the tips and for the e-beam lithographed samples. They also would like to acknowledge the help of E. Leary for reading the manuscript. A.C-G. acknowledges fellowship support from the Comunidad de Madrid (Spain). This work was supported by MICINN (Spain) (MAT2008-01735 and CONSOLIDER en Nanociencia molecular CSD-2007-00010) and Comunidad de Madrid (Spain) through the program Citecnomik (S_0505/ESP/0337).

### References

[1] Binnig G and Rohrer H 1982 Scanning Tunneling Microscopy *Helvetica Physica Acta* **55** 726-35
[2] Binnig G, Quate C F and Gerber C 1986 Atomic Force Microscope *Phys. Rev. Lett.* **56** 930-3
[3] Gorbunov A A, Wolf B and Edelmann J 1993 The use of silver tips in scanning tunneling microscopy *Rev. Sci. Instrum.* **64** 2393-4
[4] Kim P, Kim J H, Jeong M S, Ko D-K, Lee J and Jeong S 2006 Efficient electrochemical etching method to fabricate sharp metallic tips for scanning probe microscopes *Rev. Sci. Instrum.* **77** 103706-5
[5] Ren B, Picardi G and Pettinger B 2004 Preparation of gold tips suitable for tip-enhanced Raman spectroscopy and light emission by electrochemical etching *Rev. Sci. Instrum.* **75** 837-41
[6] Albonetti C, Bergenti I, Cavallini M, Dediu V, Massi M, Moulin J-F and Biscarini F 2002 Electrochemical preparation of cobalt tips for scanning tunneling microscopy *Rev. Sci. Instrum.* **73** 4254-6
[7] Uehara Y, Fujita T, Iwami M and Ushioda S 2001 Superconducting niobium tip for scanning tunneling microscope light emission spectroscopy *Rev. Sci. Instrum.* **72** 2097-9
[8] Xu M, Takano Y, Hatano T, Kitahara M and Fujita D 2003 The fabrication of MgB2 superconducting STM tips *Physica C: Superconductivity and its applications* **388** 117-8
[9] Edwards H, Taylor L, Duncan W and Melmed A J 1997 Fast, high-resolution atomic force microscopy using a quartz tuning fork as actuator and sensor *J. Appl. Phys.* **82** 980-4

[10]   Giessibl F J 1998 High-speed force sensor for force microscopy and profilometry utilizing a quartz tuning fork *Appl. Phys. Lett.* **73** 3956-8

[11]   Edwards H, Taylor L, Duncan W and Melmed A 1997 Fast, high-resolution atomic force microscopy using a quartz tuning fork as actuator and sensor *J. Appl. Phys.* **82** 980

[12]   Hembacher S, Giessibl F J, Mannhart J and Quate C F 2005 Local spectroscopy and atomic imaging of tunneling current, forces, and dissipation on graphite *Phys. Rev. Lett.* **94**

[13]   Rubio-Bollinger G, Joyez P and Agrait N 2004 Metallic adhesion in atomic-size junctions *Phys. Rev. Lett.* **93** 116803

[14]   Valkering A, Mares A, Untiedt C, Gavan K, Oosterkamp T and van Ruitenbeek J 2005 A force sensor for atomic point contacts *Rev. Sci. Instrum.* **76** 103903

[15]   Castellanos-Gomez A, Agraït N and Rubio-Bollinger G 2009 Dynamics of quartz tuning fork force sensors used in scanning probe microscopy *Nanotechnology* **20** 8pp

[16]   Naber A 1999 The tuning fork as sensor for dynamic force distance control in scanning near-field optical microscopy *Journal of Microscopy* **194** 307-10

[17]   Ng B, Zhang Y, Wei Kok S and Chai Soh Y 2009 Improve performance of scanning probe microscopy by balancing tuning fork prongs *Ultramicroscopy* **109** 291-5

[18]   Rychen J 2001 Combined low-temperature scanning probe microscopy and magneto-transport experiments for the local investigation of mesoscopic systems. Doktorarbeit, Swiss Federal Institute of Technology ETH Zürich)

[19]   Ohmori T, Nagahara L A, Hashimoto K and Fujishima A 1994 Characterization of carbon material as a scanning tunneling microscopy tip for in situ electrochemical studies *Rev. Sci. Instrum.* **65** 404-6

[20]   Colton R, Baker S, Baldeschwieler J and Kaiser W 1987 ''Oxide-free''tip for scanning tunneling microscopy *Appl. Phys. Lett.* **51** 305

[21]   Konishi H, Murata Y, Wongwiriyapan W, Kishida M, Tomita K, Motoyoshi K, Honda S, Katayama M, Yoshimoto S, Kubo K, Hobara R, Matsuda I, Hasegawa S, Yoshimura M, Lee J G and Mori H 2007 High-yield synthesis of conductive carbon nanotube tips for multiprobe scanning tunneling microscope *Rev. Sci. Instrum.* **78** 013703-6

[22]   Nguyen C, Chao K, Stevens R, Delzeit L, Cassell A, Han J and Meyyappan M 2001 Carbon nanotube tip probes: stability and lateral resolution in scanning probe microscopy and application to surface science in semiconductors *Nanotechnology* **12** 363-7

[23]   Uchihashi T, Higgins M, Nakayama Y, Sader J and Jarvis S 2005 Quantitative measurement of solvation shells using frequency modulated atomic force microscopy *Nanotechnology* **16** 49

[24]   Yeong K, Boothroyd C and Thong J 2006 The growth mechanism and field-emission properties of single carbon nanotips *Nanot* **17** 3655-61

[25]   Arai T, Gritschneder S, Troger L and Reichling M 2004 Carbon tips as sensitive detectors for nanoscale surface and sub-surface charge *Nanot* **15** 1302-6

[26]   Wong S, Woolley A, Joselevich E, Cheung C and Lieber C 1998 Covalently-Functionalized Single-Walled Carbon Nanotube Probe Tips for Chemical Force Microscopy *J. Am. Chem. Soc* **120** 8557-8

[27]   Kahng Y, Choi J, Park B, Kim D, Choi J, Lyou J and Ahn S 2008 The role of an amorphous carbon layer on a multi-wall carbon nanotube attached atomic force microscope tip in making good electrical contact to a gold electrode *Nanotechnology* **19** 195705

[28]   Buchoux J, Aimé J, Boisgard R, Nguyen C, Buchaillot L and Marsaudon S 2009 Investigation of the carbon nanotube AFM tip contacts: free sliding versus pinned contact *Nanotechnology* **20** 475701

[29]   Wei H, Kim S, Zhao M, Ju S, Huey B, Marcus H and Papadimitrakopoulos F 2008 Control of length and spatial functionality of single-wall carbon nanotube AFM nanoprobes *Chem. Mater.* **20** 2793-801

[30]   Zhao M, Sharma V, Wei H, Birge R, Stuart J, Papadimitrakopoulos F and Huey B 2008 Ultrasharp and high aspect ratio carbon nanotube atomic force microscopy probes for enhanced surface potential imaging *Nanotechnology* **19** 235704

[31]   Ohmori T, Nagahara L, Hashimoto K and Fujishima A 1994 Characterization of carbon material as a scanning tunneling microscopy tip for in situ electrochemical studies *Rev. Sci. Instrum.* **65** 404

[32]   Smit R H, Grande R, Lasanta B, Riquelme J J, Rubio-Bollinger G and Agrait N 2007 A low temperature scanning tunneling microscope for electronic and force spectroscopy *Rev Sci Instrum* **78** 113705

[33]   Berdunov N, Pollard A and Beton P 2009 Dynamic scanning probe microscopy of adsorbed molecules on graphite *Appl. Phys. Lett.* **94** 043110

[34]   Yao J, Yu W and Pan D 2008 Tensile strength and its variation of PAN-based carbon fibers. III. Weak-link analysis *J. Appl. Polym. Sci.* **110** 3778-84

[35]   Johnson D 1987 Structure-property relationships in carbon fibres *J. Phys. D: Appl. Phys.* **20** 286-91

[36]   Albrecht T, Grütter P, Horne D and Rugar D 1991 Frequency modulation detection using high-Q cantilevers for enhanced force microscope sensitivity *J. Appl. Phys.* **69** 668

[37]   Giessibl F 2001 A direct method to calculate tip–sample forces from frequency shifts in frequency-modulation atomic force microscopy *Appl. Phys. Lett.* **78** 123

[38]   Dürig U, Gimzewski J and Pohl D 1986 Experimental observation of forces acting during scanning tunneling microscopy *Phys. Rev. Lett.* **57** 2403-6

[39]   Oral A, Grimble R, Özer H, Hoffmann P and Pethica J 2001 Quantitative atom-resolved force gradient imaging using noncontact atomic force microscopy *Appl. Phys. Lett.* **79** 1915

[40]   Meepagala S and Real F 1994 Detailed experimental investigation of the barrier-height lowering and the tip-sample force gradient during STM operation in air *Phys. Rev. B* **49** 10761-3

[41]   Hahn J, Hong Y and Kang H 1998 Electron tunneling across an interfacial water layer inside an STM junction: tunneling distance, barrier height and water polarization effect *Applied Physics A: Materials Science & Processing* **66** 467-72

[42]     Hong Y, Hahn J and Kang H 1998 Electron transfer through interfacial water layer studied by scanning tunneling microscopy *The Journal of Chemical Physics* **108** 4367
[43]     Denk W and Pohl D 1991 Local electrical dissipation imaged by scanning force microscopy *Appl. Phys. Lett.* **59** 2171
[44]     Stowe T, Kenny T, Thomson D and Rugar D 1999 Silicon dopant imaging by dissipation force microscopy *Appl. Phys. Lett.* **75** 2785
[45]     Chen C 1988 Theory of scanning tunneling spectroscopy *J. Vac. Sci. Technol. A* **6** 319
[46]     Klusek Z 1998 Scanning tunneling microscopy and spectroscopy of the thermally oxidized (0001) basal plane of highly oriented pyrolitic graphite *Appl. Surf. Sci.* **125** 339-50
[47]     Glatzel T, Sadewasser S and Lux-Steiner M 2003 Amplitude or frequency modulation-detection in Kelvin probe force microscopy *Appl. Surf. Sci.* **210** 84-9
[48]     Yu Y J, Zhao Y, Ryu S, Brus L E, Kim K S and Kim P Tuning the Graphene Work Function by Electric Field Effect *Nano Lett.* 351-5
[49]     Ouseph P, Poothackanal T and Mathew G 1995 Honeycomb and other anomalous surface pictures of graphite *Phys. Lett. A* **205** 65-71
[50]     Paredes J, Mart nez-Alonso A and Tascon J 2001 Triangular versus honeycomb structure in atomic-resolution STM images of graphite *Carbon* **39** 476-9
[51]     Tománek D, Louie S, Mamin H, Abraham D, Thomson R, Ganz E and Clarke J 1987 Theory and observation of highly asymmetric atomic structure in scanning-tunneling-microscopy images of graphite *Phys. Rev. B* **35** 7790-3
[52]     Rohlfing D and Kuhn A 2007 Scanning Tunneling Microscopy of Electrode Surfaces Using Carbon Composite Tips *Electroanalysis* **19**
[53]     Family F 1999 Scaling, percolation and coarsening in epitaxial thin film growth *Physica A* **266** 173-85
[54]     Cruz T G S, Kleinke M U and Gorenstein A 2002 Evidence of local and global scaling regimes in thin films deposited by sputtering: An atomic force microscopy and electrochemical study *Appl. Phys. Lett.* **81** 4922
[55]     Sánchez F, Infante I C, Lüders U, Abad L and Fontcuberta J 2006 Surface roughening by anisotropic adatom kinetics in epitaxial growth of La0. 67Ca0. 33MnO3 *Surf. Sci.* **600** 1231-9
[56]     Rychen J, Ihn T, Studerus P, Herrmann A and Ensslin K 1999 A low-temperature dynamic mode scanning force microscope operating in high magnetic fields *Rev. Sci. Instrum.* **70** 2765
[57]     Rychen J, Ihn T, Studerus P, Herrmann A, Ensslin K, Hug H, van Schendel P and Güntherodt H 2000 Operation characteristics of piezoelectric quartz tuning forks in high magnetic fields at liquid helium temperatures *Rev. Sci. Instrum.* **71** 1695